\documentclass[aps,prl,twocolumn,showpacs,superscriptaddress,groupedaddress]{revtex4}  % for review and submission
\usepackage{graphicx}  % needed for figures
\usepackage{dcolumn}   % needed for some tables
\usepackage{bm}        % for math
\usepackage{amssymb}   % for math
\usepackage{amsmath}   % for math
\makeatletter
\newcommand*{\rom}[1]{\expandafter\@slowromancap\romannumeral #1@}
\makeatother

\begin{document}

\title{Analytical Solutions of the Dirac and the Klein-Gordon Equations in Plasma Induced by High Intensity Laser}
\author{Erez Raicher\footnote{E-mail address: erez.raicher@mail.huji.ac.il }}
\affiliation{Racah Institute of Physics, Hebrew University, Jerusalem 91904, Israel }
\affiliation{Department of Applied Physics, Soreq Nuclear Research Center, Yavne 81800, Israel }
\author{Shalom Eliezer}
\affiliation{Department of Applied Physics, Soreq Nuclear Research Center, Yavne 81800, Israel }
\affiliation{Nuclear Fusion Institute, Polytechnic University of Madrid, Madrid, Spain }
%\input author_list.tex       % D0 authors (remove the first 3 lines
%                             % of this file prior to submission, they
%                             % contain a time stamp for the authorlist)
%                             % (includes institutions and visitors)
\date{\today}

\begin{abstract}
In this paper we obtain analytical solutions of the Dirac and the Klein-Gordon equations coupled to a strong electromagnetic wave in the presence of plasma environment. These are a generalization of the familiar Volkov solutions. The contribution of the non-zero photon effective mass to the scalar and fermion wavefunctions, conserved quantities and effective mass is demonstrated for the first time. The new wavefunctions exhibit differences from Volkov solutions for nowadays available laser intensity. 
 \end{abstract}

\maketitle

%\section{\label{sec:level1}First-level heading}
% sections are not used for PRL papers
\section{\rom{1}. Introduction}
In the near future ultraintense lasers with irradiances $I_L \sim 10^{25}  W/cm^2$ are expected to be available \cite {ELI,XCELS,HIPER,Tajima, DiPiazza}. Such intensities will allow laboratory exploration of plethora of physical phenomena, among which QED in strong fields \cite{DiPiazza, mendonca2}, Schwinger vacuum decay \cite{schwinger} and Unruh radiation \cite{unruh1, unruh2}. The above is additional to the conventional applications of ultraintense lasers: fast ignition \cite{FI}, ions acceleration \cite{ion acc}, high harmonics generation \cite{atto} and relativisitc shock waves \cite{shock}. Our understanding of these phenomena relies on the basic theoretical description of the interaction of an electromagnetic field with an electron. This interaction does not obey classical electrodynamics if the electron proper acceleration in its rest frame is comparable to the Schwinger acceleration $a_S \equiv mc^3/h$, where $m$ is the electron mass, $c$ is the speed of light and h is Planck constant. An electron experiencing Schwinger acceleration gains, by definition, an energy equal to its rest mass over a distance of one Compton wavelength $\lambda_C =h / mc $.  Equivalently, the dynamics has quantum nature if 
\begin{equation}
\chi \equiv \frac{h}{mc^3} \sqrt{\left ( \frac{du}{ds} \right ) ^2}>1
\end{equation}
where $s$ is the proper time of the electron and  $u^{\mu}$ is the proper velocity. Moreover, the motion is nonlinear in the electromagnetic field amplitude if 
\begin{equation}
\xi \equiv \frac{e}{m} \sqrt{-A^2} >1
\end{equation}
where $A_{\mu}$ is the vector potential and e is the electric charge. In practical units the normalized vector potential is given by $\xi^2 = \lambda_L^2 I_L / (1.37 \times 10^{18}W \mu m^2 / cm^2)$ where $\lambda_L$ is the laser wavelength. For the laser intensities mentioned above both conditions are satisfied, calling for a nonperturbative QED formalism. The essence of the nonperturbative attitude (also known as "Furry picture" \cite{furry}) is that instead of treating the laser background perturbatively, we include it in the free Hamiltonian. Therefore, nonperturbative calculations of QED processes in the presence of laser ("laser assisted") are carried out \cite{QED1, QED2, QED3, QED4, QED5} by replacing the free electron wavefunction appearing in the quantum calculation with the familiar Volkov solution \cite{Volkov} for an electron interacting with an electromagnetic plane wave. 

Laser assisted QED processes in the nonperturbative regime involve absorption of many laser photons. The most dominant processes are the nonlinear Compton scattering, where an electron absorbs many laser photons and emits a gamma photon, and the nonlinear Breit-Wheeler scattering which involves a gamma photon decaying into an electron-positron pair under the influence of the laser field \cite{DiPiazza}. These two processes are responsible for the QED cascades expected to play a key role \cite{elkina} in the dynamics of the plasma created during the laser-matter interaction. In recent years, a considerable effort is invested in order to extend the nonperturbative attitude to realistic laser configurations, e.g. by taking into account the finite temporal width of the laser pulse \cite{finitePulse, finitePulse2, finitePulse3}. Additional issue to be treated is the plasma effect on the laser dispersion relation. This problem was considered  in \cite{mendonca} under low density assumption.

In this work, we consider the problem of a particle in the presence of an electromagnetic wave propagating in plasma environment. We start with a scalar charged particle described by the Klein-Gordon equation, due to its relative simplicity. It will serve as a reference to the Dirac equation, for which we derive an approximate solution utilizing Floquet theorem and solid state analogy. The new wavefunction can be substituted in the perturbative procedure in order to calculate the electron energy loss due to photon emission as well as other QED processes. The new solutions are of high relevance if the laser interacts with dense matter or in the case of counter propagating beams in vacuum (i.e. a rotating electric field).

The paper is organized as follows. Section \rom{2} includes the exact solution of the Klein-Gordon equation in the presence electromagnetic field in plasma environment. In section \rom{3} we describe in details the approximate solution of the Dirac equation for the same problem. In section \rom{4} we obtain the conserved quantities of the problem and the particle effective mass. Finally, we discuss our results and conclude in section \rom{5}.

\section{\rom{2}. Klein-Gordon Equation Solution}
Let us start with the Klein-Gordon equation for a charged scalar with mass $M$ coupled to a classical electromagnetic field 
\begin{equation}
\left [ (i {\partial}  -eA)^2  -M^2 \right] \Phi = 0
\label{eq:av Dirac1}
\end{equation}
From now on natural units ($\hbar=c=1$) are used. Notice that $\partial \cdot (A \Phi) = A \cdot \partial \Phi + (\partial \cdot A) \Phi $. We adopt the Lorentz gauge ($\partial \cdot A = 0$) and therefore the last term vanishes.
\begin{equation}
\left[-\partial^2-2ie(A\cdot\partial)+e^2A^2-M^2 \right]\Phi = 0
\label{eq:av KG3}
\end{equation}
The dot symbol stands for Lorentz contraction. $ k \equiv (\omega_L,0,0,k_z)$ is the wavenumber of the laser and the vector potential $A_{\mu}$ depends only on the quantity $\phi \equiv k \cdot x$. 
We are interested in a laser wave propagating through dense plasma. It is well known that in plasma environment the photon acquires effective mass due to the screening \cite{kapusta}. The general dispersion relation is
\begin{equation}
k^2 = \omega_L^2-\vec{\textbf{k}}^2 = m_{ph}^2
\label{eq:av dispersion}
\end{equation} 
The effective mass $m_{ph}$ will be treated later in this paper. Unless we are dealing with a few cycle pulse, it is acceptable to assume a plane monochromatic wave, so that the vector potential may be written as
\begin{equation}
A(\phi) = a(\phi) \left ( \epsilon e^{i\phi} +  {\epsilon}^* e^{-i\phi} \right )
\label{eq:av Apot}
\end{equation}
where $a(\phi)$ is an envelope that slowly goes to zero as $\phi \rightarrow \pm \infty$. From now on we treat $a(\phi)$ as a constant, similiarly to the standard Volkov derivation \cite{landau}.  If the polarization is circular, we have 
\begin{equation}
\epsilon =( e_{1}-ie_{2} ) / 2 
\end{equation}
 where
the unit vectors are $e_{1} = (0,1,0,0)$ and $e_{2} = (0,0,1,0)$.
The polarization vector obeys ${\epsilon}^2 = {\epsilon^*}^2 = 0$ as well as $\epsilon \cdot  \epsilon^*= - \frac{1}{2}$. It can be easily verified that $A^2 = -a^2 $.

Before we dive into the derivation, let us recall that in the interaction of a single particle with an electromagnetic wave the quantum processes are governed by 4 parameters \cite{ritus}: $\xi$ and $\chi$ defined above as well as the electromagnetic field invariants 
\begin{equation}
\mathcal{F} \equiv -\frac{1}{4}F_{\mu \nu}F^{\mu \nu}
\end{equation}
 and 
\begin{equation}
\mathcal{G} =\frac{1}{4}\epsilon_{\alpha \beta \mu \nu} F^{\alpha \beta}F^{\mu \nu} 
\end{equation}
Einstein convention is used to summarize over identical greek indices from 0 to 3.
The electromagnetic field tensor is related to the vector potential by $F_{\mu \nu} \equiv {\partial}_\mu A_{\nu} - {\partial}_\nu A_{\mu} $ and $\epsilon_{\alpha \beta \mu \nu}$ is the Levi Civita tensor. In terms of the electric field $\textbf{E}$ and the magnetic field \textbf{B} it acquires the familiar form
\begin{equation}
\mathcal{F} =- \frac{1}{2}(\textbf{E}^2 - \textbf{B}^2) \\
\end{equation}
\begin{equation}
\mathcal{G} =- \textbf{E} \cdot \textbf{B} 
\end{equation}
 If the electromagnetic wave is propagating in vacuum, both invariants are zero. However, if plasma is present then $\mathcal{G}$ remains zero and $\mathcal{F}  = -a^2 {m_{ph}}^2 /2 $. Hence, the QED cross sections obtained with the solutions derived in this paper will depend upon additional independent variable comparing to the Volkov solutions which ignore the plasma effect. 

In the following derivation, the effective photon mass $m_{ph}$ can be either calculated self consistently (as will be published in a separate paper) or be taken as an external input from a kinetic model, such as a particle-in-cell (PIC) code. Nowadays, PIC simulations for ultraintense laser plasma interaction model QED processes (such as nonlinear Compton scattering and nonlinear Breit-Wheeler) with rates calculated using Volkov wavefunctions \cite{elkina}. As a result, they depend only on the parameters $\chi$ and $\xi$ evaluated at the particle position. The solutions exhibited here can be utilized to rates calculation depending on an additional parameter, $\mathcal{F}=-a^2 m_{ph}^2 /2$.
 
Since $A_{\mu}$ depends only on $\phi$, we seek a solution for (\ref{eq:av KG3})  in the form
\begin{equation}
\Phi = e^{-ipx}G(\phi)
\label{eq:av guess_KG}
\end{equation}
$p_{\mu}$ is a constant 4-vector which is the momentum of the particle in the limit $k \cdot x \rightarrow \pm \infty$ where the electromagnetic field is absent. Hence we have $p^2=M^2$.
Substituting (\ref{eq:av guess_KG}) into (\ref{eq:av KG3}) we get a $2^{nd}$ order linear ordinary differential equation
\begin{multline}
-m_{ph}^2G''+2i(k\cdot p)G'+ 
\left[-2e(p\cdot A) + e^2A^2 
  \right]G=0
\label{eq:av KG}
\end{multline}
where the tag symbol represents the derivative with respect to $\phi$. For a plane wave traveling in vacuum we have $m_{ph}^2 = 0$, so that we are left with a $1^{st}$ order equation. Its solution is the familiar Volkov wavefunction
\begin{equation}
\Phi = e^{iS} 
\label{eq:av Volkov_KG}
\end{equation}
where S is the classical action of a particle in electromagnetic field
\begin{equation}
S \equiv -p \cdot x - \int^{\phi}_{0}  \left [ \frac{e}{k \cdot p} (p \cdot A) - \frac{e^2}{2k \cdot p} A^2 \right ] d \phi'
\label{eq:av action}
\end{equation}
We now assume $A \cdot p = 0$ since the plasma is initially at rest, i.e. the particles velocities are nonrelativistic leading to $p \approx (M,0,0,0)$. This condition also holds for a particles beam counter propagating with respect to the laser beam. Therefore we have
\begin{equation}
-m_{ph}^2G''+2i(k\cdot p)G'
 - e^2a^2 G=0
\label{eq:av KG4}
\end{equation}
The resultant equation has constant coefficients and hence its solution is trivial:  $G = e^{i \nu \phi}$. We substitute it into (\ref{eq:av KG4}) and get an equation for $\nu$.
\begin{equation}
{m_{ph}}^2 \nu^2 - 2 (k \cdot p) \nu -e^2a^2 = 0
\end{equation}
So that $\nu$ is given by
\begin{equation}
\nu = \frac{ k \cdot p  }{ {m_{ph}}^2} \left ( 1 -  \sqrt{1 + \left (  \frac{ ea {m_{ph}} }{ k \cdot p } \right )^2 }  \right )
\label{eq:av nu_kg}
\end{equation}
The second solution of the quadratical equation is not physical because it does not go to 0 in the limit $ k \cdot x \rightarrow \pm \infty $. The Final solution is
\begin{equation}
\Phi = e^{-i(p-\nu k) \cdot x} 
\label{eq:av KG_f}
\end{equation}
In the next section we show that the solution of the Dirac equation yields the same value for $\nu$.

\section{\rom{3}. Dirac Equation Solution}
Having examined the simple case of Klein-Gordon, we proceed to the Dirac equation for an electron coupled to a classical electromagnetic field
\begin{equation}
\left [i {\not}{\partial}  -e{\not}A  -m \right] \psi = 0
\label{eq:av Dirac1}
\end{equation}
${\not}A $ stands for $ \gamma_{\mu}A^{\mu} = \gamma \cdot A$, where $\gamma_{\mu}$ are the Dirac matrices and $\psi$ is the four-component wavefunction. Now we transform to the $2^{nd}$ order Dirac equation by multiplying (\ref{eq:av Dirac1}) with the operator $i {\not}{\partial}  -e{\not}A  +m$.
\begin{equation}
\left[-\partial^2-2ie(A\cdot\partial)+e^2A^2-m^2-ie{\not} k {\not} {A'} \right]\psi_s = 0
\label{eq:av Dirac3}
\end{equation}
The 's' subscript was added to $\psi$ to emphasize that it solves the $2^{nd}$ order Dirac equation. The relation between $\psi$ and $\psi_s$ will be discussed in the end of this section.  The vector potential and the dispersion relation are given by (\ref{eq:av Apot}) and (\ref{eq:av dispersion}) correspondingly. 
Similiarly to the previous section, we seek a solution for (\ref{eq:av Dirac3})  in the form
\begin{equation}
\psi_s = e^{-ipx}F(\phi)
\label{eq:av guess}
\end{equation}
$p_{\mu}$ is a constant 4-vector which is the momentum of the electron in the limit $k \cdot x \rightarrow \pm \infty$ where the electromagnetic field is absent. Hence we have $p^2=m^2$.
Substituting (\ref{eq:av guess}) into (\ref{eq:av Dirac3}) we get a $2^{nd}$ order linear ordinary differential equation
\begin{multline}
-m_{ph}^2 F''+2i(k\cdot p)F'+ \\ 
\left[-2e(p\cdot A) + e^2A^2 
 -ie{\not} k {\not} A' \right]F=0
\label{eq:av Dirac4}
\end{multline}
For a plane wave traveling in vacuum we have $m_{ph}^2 = 0$, so that we are left with a $1^{st}$ order equation. Its solution is the familiar Volkov wavefunction
\begin{equation}
\psi = \left (1+ \frac{e}{2k \cdot p}  {\not}{k} {\not} {A} \right ) e^{iS}  u_p
\label{eq:av Volkov}
\end{equation}
where the classical action S is given in (\ref{eq:av action}).
Substituting the explicit expression for the vector potential (\ref{eq:av Apot}) into equation (\ref{eq:av Dirac4}) we get
\begin{equation}
-{m_{ph}}^2F'' + 2ie(k\cdot p)F' + \left(-e^2a^2 + \Xi e^{i\phi} + \zeta e^{-i\phi} \right)F=0
\label{eq:av Dirac5}
\end{equation}
where we have defined 
\begin{equation}
\Xi =  ae{\not}k{\not} \epsilon - 2ea (p \cdot \epsilon)
\end{equation}
 and 
\begin{equation}
\zeta = - ae{\not}k{\not} \epsilon^*- 2ea (p \cdot \epsilon^*)
\end{equation}
Note that $\Xi , \zeta $ are $4 \times 4$ matrices while the other coefficients in equation (\ref{eq:av Dirac5}) are scalars. As in section \rom{2} we assume $A \cdot p = 0$ since the plasma is initially at rest, i.e. the electrons velocities are nonrelativistic leading to $p \approx (m,0,0,0)$. This condition also holds for an electrons beam counter propagating with respect to the laser beam. Using the Lorentz gauge ($k \cdot A = 0$) one can prove that $\Xi ^2=  \zeta ^2 = 0$. Similiarly ,
\begin{equation}
 \Xi \zeta =- e^2 a^2 {m_{ph}}^2 D_1 
\label{eq:av xi_zeta}
\end{equation}
 and 
\begin{equation}
\zeta \Xi =- e^2 a^2 {m_{ph}}^2 D_2
\label{eq:av zeta_xi}
\end{equation}
where
\begin{equation}
 D_1 \equiv \left( \begin{array}{cccccc}
1& 0  &0  &0  \\
0 & 0  &0  &0  \\
0 & 0  &1  &0  \\
0 & 0  &0  &0  \\
\end{array} \right) ,  D_2 \equiv \left( \begin{array}{cccccc}
0& 0  &0  &0  \\
0 & 1  &0  &0  \\
0 & 0  &0  &0  \\
0 & 0  &0  &1  \\
\end{array} \right)
\label{eq:av D_def}
\end{equation}

Equation (\ref{eq:av Dirac5}) can be solved in a similiar way to the Mathieu equation.
The coefficients are periodic in $\phi$ and therefore we can utilize Floquet theorem
\begin{equation}
F = P(\phi)e^{i\nu\phi}
\label{eq:av bloch}
\end{equation}
where $P$ is some periodic function of $\phi$. Plugging (\ref{eq:av bloch}) into (\ref{eq:av Dirac5}) we see that $P$ obeys the equation
\begin{equation}
-{m_{ph}}^2P'' + \mu P' +\left[\delta+ \Xi e^{i\phi} + \zeta e^{-i\phi} \right]P=0
\label{eq:av Peq}
\end{equation}
The scalars $\mu$ and $\delta$ are defined as 
\begin{equation}
\mu \equiv  2i \left (  k \cdot p  -\nu{m_{ph}}^2 \right )
\end{equation}
 and 
\begin{equation}
\delta \equiv {m_{ph}}^2 \nu^2-2 \nu k \cdot p -e^2a^2 
\end{equation}
Since $P$ is periodic we write it as a Fourier series
\begin{equation}
P =  \sum_{n=-\infty}^{\infty}\eta_ne^{in\phi} 
\label{eq:av series}
\end{equation}
where $\eta_n$ are constant bispinors. Subsituting (\ref{eq:av series}) into (\ref{eq:av Peq}), the following recursive relations can be achieved
\begin{equation}
{\rho_n}\eta_n + \Xi \eta_{n-1} + \zeta\eta_{n+1}= 0
\label{eq:av Rho}
\end{equation}
with ${\rho_n}  \equiv  {m_{ph}}^2n^2 +i\mu n + \delta$.
Substituting the definitions of $\mu, \delta$, we have
\begin{equation}
{\rho_n}   = {m_{ph}}^2 (\nu + n)^2 - 2 (k \cdot p) (\nu+n) -e^2a^2
\label{eq:av rho}
\end{equation}
Taking the sum in equation (\ref{eq:av series}) from $-N$ to $N$ (for $N \gg 1$), the finite set of equations is described by the matrix equation:
\begin{equation}
\Lambda \eta = 0
\label{eq:av matrix_eq}
\end{equation}
where we have defined
\begin{equation}
{\Lambda} \equiv \left( \begin{array}{ccccc}
{\rho_{-N}}I_4 & \zeta  & &&0 \\
\Xi & {\rho_{-N+1}}I_4 & \zeta  & & \\
 & \ddots & \ddots & \ddots & \\
0&&&\Xi & {\rho_{N}}I_4   \end{array} \right)
\label{eq:av etaN}
\end{equation}
and
\begin{equation}
{\eta} \equiv \left( \begin{array}{cccccc}
\eta_{-N} \\
\vdots  \\
\eta_N   \end{array} \right)
\label{eq:av etaN}
\end{equation}
$I_4$ is the $4 \times 4$ unit matrix and the dimensions of $\Lambda$ are $4 \times (2N+1)$.
It is interesting to point out that the procedure described in equations (\ref{eq:av bloch}) - (\ref{eq:av etaN}) is similiar to the band structure calculation in solid state problems, where the vector potential is analogous to the periodic crystal potential and $\nu k_{\mu}$ resembles the quasi momentum. 

For a given value of $\nu$, at most one of the coefficients $\rho_i$ can vanish, i.e, $\rho_i \not= 0$ for $i \not= l$ where $l$ is some unknown index.
Since $\nu$ is defined up to an integer, the coefficients indices can be shifted by any integer number. Hence, we can choose $l =0$ without loss of generality. In order to calculate the bispinors $\eta_i$, we would like to express them in terms of $\eta_0$. Let us look at the equation corresponding to the first row of the matrix equation (\ref{eq:av matrix_eq}).
Since $\rho_{-N}$ is diagonal, $\eta_{-N}$ can be expressed in terms of $\eta_{-N+1}$.
\begin{equation}
\eta_{-N}= -{\rho_{-N}}^{-1}\zeta \eta_{-N+1}
\label{eq:av eta_N}
\end{equation}
Substituting (\ref{eq:av eta_N}) into the second row of equation (\ref{eq:av matrix_eq}) yields
\begin{equation}
 \eta_{-N+1}  =- \left (  -\frac{1}{\rho_{-N}}\Xi \zeta + I_4 \rho_{-N+1}     \right )^{-1}  \zeta \eta_{-N+2} 
\label{eq:av eta_N-1}
\end{equation}
We now use the following matrix identity (where $\alpha_1$ and $\alpha_2$ are constant scalars and $D_1$ is given in (\ref{eq:av D_def}))
\begin{equation}
\left (\alpha_1 I_4 + \alpha_2 D_1 \right )^{-1} = \frac{1}{\alpha_1}I_4 - \frac{\alpha_2}{\alpha_1 (\alpha_1 + \alpha_2)} D_1
\label{eq:av identity}
\end{equation}
and the relations (\ref{eq:av xi_zeta}) and $\zeta^2 = 0$ in order to obtain
\begin{equation}
 \eta_{-N+1}  =-  \frac{1}{ \rho_{-N+1}}  \zeta \eta_{-N+2} 
\label{eq:av E}
\end{equation}
Following this procedure we get in the general case
\begin{equation}
 \eta_{-i}  =-  \frac{1}{ \rho_{-i}}  \zeta \eta_{-i+1} 
\label{eq:av recursion1}
\end{equation}
For $n> 0 $ we use the same method beginning at $n = N$ and going downward. It is possible because identity (\ref{eq:av identity}) is satisfied also for $D_2.$
\begin{equation}
 \eta_{i}  =-  \frac{1}{ \rho_{i}}  \Xi \eta_{i-1} 
\label{eq:av recursion2}
\end{equation}
Due to the relation $\Xi^2 = \zeta^2  =0$ we get a truncation $ \eta_{i}  = 0$ for $i \neq 0,\pm1$. 
We are left with 
\begin{equation}
\left( \begin{array}{cccccc}
{\rho_{-1}}I_4 & \zeta  & 0 \\
\Xi & {\rho_{0}}I_4 & \zeta\\
0 & \Xi & {\rho_{1}}I_4 \end{array} \right) \left( \begin{array}{c}
\eta_{-1} \\
\eta_{0}  \\
\eta_1   \end{array} \right) = 0
\label{eq:av lambda2}
\end{equation}
If we express $\eta_1, \eta_{-1}$ in terms of $\eta_0$ we get
\begin{equation}
 \left( \rho_0  I_4 -\frac{1}{ \rho_{-1}} \Xi \zeta -  \frac{1}{\rho_{1}} \zeta \Xi \right) \eta_0 = 0
\end{equation}
In order to have a nontrivial solution, we require that the determinant vanishes, i.e.
\begin{equation}
det \left[ \rho_0  I_4 + e^2a^2 m_{ph}^2 \left(  \frac{1}{ \rho_{-1}} D_1 +  \frac{1}{\rho_{1}}D_2 \right) \right]  = 0
\label{eq:av det3}
\end{equation}
where the relations (\ref{eq:av xi_zeta}, \ref{eq:av zeta_xi}) were used. The solution of the above equation gives us $\nu$. Regarding the structure of $D_1,D_2$, equation (\ref{eq:av det3}) reduces to 2 scalar equations:
\begin{equation}
 \rho_0 \rho_{1} + e^2a^2 {m_{ph}}^2  = 0
\label{eq:av det4}
\end{equation}
\begin{equation}
 \rho_0 \rho_{-1} + e^2a^2 {m_{ph}}^2  = 0
\label{eq:av det2}
\end{equation}
We assume $e^2a^2 \gg {m_{ph}}^2$, or equivalently $(\xi m / m_{ph})^2 \gg 1$ which is a very good approximation for optical high intensity laser ($m_{ph} < \omega \approx 1eV$). Therefore, the second term in (\ref{eq:av det3}) is negligible with respect to the constant part of $\rho_{1} \rho_0$. Consequently, equation (\ref{eq:av det4}) reduces to $\rho_0 = 0$. Similiar argument holds for (\ref{eq:av det2}).
It allows us to find $\nu$
\begin{equation}
\nu = \frac{ k \cdot p  }{ {m_{ph}}^2} \left ( 1 -  \sqrt{1 + \left (  \frac{ ea {m_{ph}} }{ k \cdot p } \right )^2 }  \right )
\label{eq:av nu_D}
\end{equation}
The formula for $\nu$ is identical to the case of Klein-Gordon (\ref{eq:av nu_kg}). The second solution of the quadratical equation $\rho_0 = 0$ is not physical because it does not go to 0 in the limit $ k \cdot x \rightarrow \pm \infty $.
Subsituting the expression (\ref{eq:av nu_D}) for $\nu$ into (\ref{eq:av rho}) we get
\begin{equation}
{{\rho_{n}}} = -2n(k \cdot p) \sqrt{ 1+ \left ( \frac{ea m_{ph}}{k \cdot p}   \right)^2 } +  {m_{ph}}^2 n^2
\end{equation}

Having attained closed formulas for $\nu$ and $\eta_i$, the solution of the $2^{nd}$ order Dirac equation, $\psi_s$, can be expressed in a Volkov-like form:
\begin{equation}
\psi_s = e^{-i (p-  \nu k) \cdot x}  \left ( 1 -  \frac{1}{\rho_1} \Xi e^{i \phi} - \frac{1}{\rho_{-1}}  \zeta e^{-i\phi}   \right ) \eta_0
\label{eq:av mySol1}
\end{equation}
Under the assumption $ea \gg  m_{ph}$, the coefficient $\rho_1$ may be approximated by 
\begin{equation}
{{\rho_{1}}}= - \rho_{-1} =  -2(k \cdot p) \sqrt{ 1+ \left ( \frac{ea m_{ph}}{k \cdot p}   \right)^2 }
\end{equation}
Accordingly, the wavefunction(\ref{eq:av mySol1}) simplifies to
\begin{equation}
\psi _s= e^{-i (p-  \nu k) \cdot x}  \left ( 1 -  \frac{e}{\rho_{1}} {\not}k {\not}A   \right ) \eta_0
\label{eq:av mySol2}
\end{equation}
However, the $2^{nd}$ order Dirac equation includes a redundant solution which does not satisfy the original Dirac equation. Fortunately, the solution of the $1^{st}$ order Dirac equation is related to the solution of the seocnd order one by the simple transformation $\psi = (i{\not}\partial -e{\not}A + m ) \psi_s$ \cite{landau}. Therefore, the final solution takes the form
\begin{multline}
\psi = \Bigl [  1- \frac{{\not}k}{2m} \left ( \nu - \frac{e^2a^2}{\rho_1} \right ) - \frac{e}{\rho_1} {\not}k  {\not}A  \\ - \frac{e}{4m} \left ( 1 + \frac{2 k \cdot p}{\rho_1} \right ) {\not}A      -  \frac{ie {m_{ph}}^2}{2m \rho_1} {\not}A'     \Bigr ]  e^{-i(p-\nu k) \cdot x} \eta_0
\label{eq:av psi_f1}
\end{multline}
Now we have to determine $\eta_0$. From equations (\ref{eq:av det2} -  \ref{eq:av det3}) we deduce that the degeneracy order of $\Lambda$ is 4 (when the term $e^2a^2m_{ph}^2$ is neglected). For this reason we have 4 degrees of freedom, meaning that $\eta_{0}$ can be arbitrarily chosen. However, for $k \cdot x \rightarrow \pm \infty$ the solution has to go to the free particle wavefunction, implying $\eta_0 =u_p$. 

For $m_{ph} = 0$ we have $\rho_{1} =- 2(k \cdot p)$ and additionally (\ref{eq:av nu_D}) reduces to $\nu = - \frac{e^2 a^2}{2 ( k \cdot p)}$.
Consequently, our solution recovers the Volkov wavefunction (see equation (\ref{eq:av Volkov})) if the photon effective mass vanishes. The only assumption we have made is $ea>> m_{ph}$ and hence our solution is applicable for all the intensity range relevant for high intensity lasers.
One can verify that our final solution (\ref{eq:av psi_f1}) with $\nu$ given by (\ref{eq:av nu_D})  solves the Dirac equation (\ref{eq:av Dirac4}).
Obviously, the expression for the wavefunction exhibits significant deviation from the Volkov expression if $(eam_{ph} / k \cdot p)^2 >1$ is satisfied. Interestingly, this result contradicts the common assumption that as long as $\mathcal{F} / E_s^2<<0$ where $E_s = m^2/e$ is the Schwinger critical field, the particle wavefunction is approximated by Volkov solution \cite{ritus, DiPiazza}. 

\section{\rom{4}. Conserved Quantities}
According to quantum mechanics, the temporal dynamics of an observable $O$ are determined by the commutation relation of its correspondent operator $\hat{O}$ with the Hamiltonian
\begin{equation}
\frac{d}{dt} \langle \hat{O} \rangle =i\langle [H,\hat{O}] \rangle + \langle \frac{\partial \hat{O}}{\partial t} \rangle
\end{equation}
The operators discussed below have no explicit time dependence, meaning that the second term is identically zero.

It is well known \cite{ritus} that in the Volkov case, the following operators $i\partial_x, i\partial_y, i(k \cdot \partial)$ commute with the Dirac Hamiltonian corresponding to equation (\ref{eq:av Dirac1}) 
\begin{equation}
H_D = \gamma_0 [ \boldsymbol{\gamma} \cdot (- i\boldsymbol{\nabla} - e\boldsymbol{A})+m] +eA_0
\end{equation}
where $\boldsymbol{\gamma}$ stands for $(\gamma_1,\gamma_2, \gamma_3)$. For the operators  $i\partial_x, i\partial_y$ this statement is straightforward since $A_{\mu}$ does not depend on $x,y$. Let us prove it explicitly for $(k \cdot \partial)$. Since $(k \cdot \partial)$ commutes with the nabla operator, the commutation relation reads
\begin{equation}
[k \cdot \partial,H_D] = [k \cdot \partial,eA_0] -  [k \cdot \partial,e \gamma_0 \boldsymbol{\gamma} \cdot \boldsymbol{A}] 
\label{eq:av commute}
\end{equation}
Consider the first term in (\ref{eq:av commute}). According to the differentiation chain rule 
\begin{equation}
(k \cdot \partial )A_0 = A_0   (k \cdot \partial)  + [(k \cdot \partial )A_0]
\label{eq:av dif1}
\end{equation}
Because of $[(k \cdot \partial) A_0] = m_{ph}^2A_0 = 0$, we get
\begin{equation}
 [(k \cdot \partial),eA_0] = e (k \cdot \partial) A_0 - eA_0   (k \cdot \partial) =0
\label{eq:av dif2}
\end{equation}
Using the same procedure one can show that the second term in (\ref{eq:av commute}) vanishes as well, and hence $[(k \cdot \partial),H_D] = 0$.

However, if $m_{ph} \neq 0$ the above derivation is no longer valid. In order to construct an alternative conserved operator, we define 
\begin{equation}
\tilde{k} \equiv (k_z,0,0,\omega_L)
\end{equation}
Since $\tilde{k} \cdot k = 0$, one can verify that the operator $\tilde{k} \cdot \partial$ does commute with $H_D$. The action of the operators $i\partial_x, i\partial_y, i\tilde{k} \cdot \partial$ on the wavefunction given by (\ref{eq:av psi_f1}) yields $p_x, p_y, \tilde{k} \cdot p$ correspondingly. Therefore, $p_{\mu}$ is a conserved quantity. As a result, the orthogonality of the wavefunctions, which are characterized by $p_{\mu}$, is assured. Notice that if $m_{ph} = 0$ then $\tilde{k} = k$ and the third conserved quantity is the same as in the Volkov case. 

The above arguments hold for the Klein-Gordon Hamiltonian as well. For comfort reasons, we work with the two-component representation of the wavefunction, i.e. $\Phi$ is replaced by $ \left( \begin{array}{cccccc}
\theta  \\
\chi  \\
\end{array} \right)$ where
\begin{equation}
\theta \equiv \frac{1}{2} \left(   \phi + \frac{i}{M} \frac{\partial \phi}{\partial t} \right)
\end{equation}
\begin{equation}
\chi \equiv \frac{1}{2} \left(   \phi - \frac{i}{M} \frac{\partial \phi}{\partial t} \right)
\end{equation}
The Hamiltonian takes the form \cite{bjorken}
\begin{multline}
H_{KG} =   \left ( \begin{array}{rr}
1& 1   \\
-1 & -1   \\
\end{array} \right)  \frac{( - i\boldsymbol{\nabla} - e\boldsymbol{A})^2}{2M} +  \\
\left( \begin{array}{rr}
1& 0   \\
0 & -1   \\
\end{array} \right)  M + eA_0
\end{multline}
Evidently, the three operators mentioned above commute with $H_{KG}$ (the proof is analogous to (\ref{eq:av dif1} - \ref{eq:av dif2})).

Now we would like to obtain expression for the effective mass of the particle in the presence of the electromagnetic field. For this purpose, we recall that due to the periodicity of the vector potential, the wavefunction may be cast into the form
\begin{equation}
\psi = e^{-i q \cdot x} \tilde{P}(\phi)
\end{equation}
where $\tilde{P}$ is some periodic function of $\phi$. $q_{\mu}$ is called the quasi momentum since it appears in the 4-momentum conservation delta functions associated with the Feynman diagrams of laser assisted QED processes \cite{ritus}.
As a result, the effective mass is $m^* \equiv \sqrt{q^2}$.
One can deduce the quasi momentum related to the wavefunction (\ref{eq:av psi_f1}) found in the previous section:
 $q_{\mu} =  p_{\mu} - \nu k_{\mu}$. Thus, the effective mass is
\begin{equation}
\frac{m^*}{m}   = \frac{1}{m}\sqrt{(p_{\mu} - \nu k_{\mu})^2}  = \sqrt {1 + \frac{e^2a^2}{m^2} }
\label{eq:av effMass1}
\end{equation}
Though $\nu$ given by (\ref{eq:av nu_D}) is not the same as in Volkov solution \cite{landau}, the formula (\ref{eq:av effMass1}) is identical to the formula obtained with Volkov wavefunction. In other words, the effective mass is not affected by the plasma presence. For a scalar particle, $\nu$ is same and therefore its effective mass is also (\ref{eq:av effMass1}) with $m \rightarrow M$.

\section{\rom{5}. Conclusions}
We have derived analytical solutions for the Volkov problem (i.e, a Klein-Gordon or a Dirac particle under the influence of an electromagnetic wave) taking into account plasma dispersion relation. After Volkov solutions \cite{Volkov}, a very long time has passed without introducing the plasma contribution to the particle wavefunction. To our opinion, this fact in itself deserves the consideration of the scientific community. The solutions are of high relevance for laser matter interaction or for a standing wave (i.e rotating electric field) created by counter propagating pulses. The assumptions in our derivation are $ea \gg m_{ph}$ as well as $p \cdot A = 0$. Therefore, it is applicable for any realistic intense laser parameters as long as the matter is initially at rest. We have shown that if $(ea m_{ph} /  k \cdot p)^2 >1$, the differences between the Volkov solutions and our solutions are significant. In the case of an electron initially at rest, this condition is equivalent to $(\xi m_{ph} / \omega_L)^2 > 1$. For optical lasers and dense plasma (i.e. $m_{ph} \sim \omega_L$), it corresponds to $I_L > 10^{18} W/cm^2$ which is already available experimentally. This result contradicts the common assumption in the literature that as long as  $\mathcal{F} / E_s^2 \ll 0$, Volkov wavefunction is valid \cite{ritus, DiPiazza}.  Moreover, the new solutions enable us to deduce the effective mass of the particle in the presence of the electromagnetic field. Interestingly, the effective mass is not affected by the plasma environment. The photon effective mass was considered in this paper as an external input (e.g. from a PIC code). An extension of the model to include the general case of $p \cdot A \neq 0$ as well as self consistent calculation of $m_{ph}$ will be published soon. To conclude, the new solutions presented in this work may pave the way towards calculation of laser assisted QED cross sections  in plasma environment.

\end{document}